# University at Stony Brook

## CEAS Technical Report Nr 831

# Cities of the Future: Employing Wireless Sensor Networks for Efficient Decision Making in Complex Environments

Alex Doboli, Daniel Curiac[†], Dan Pescaru[‡], Simona Doboli[*], Wendy Tang,
Costantin Volosencu[†], Michael Gilberti, Ovidiu Banias[†,] Codruta Istin[‡]

Department of Electrical and Computer Engineering
State University of New York at Stony Brook, Stony Brook, NY 11794-2350

[†] Department of Automatics and Technical Informatics
"Politehnica" University Timisoara, Timisoara, Romania

[‡] Department of Computer Science
"Politehnica" University Timisoara, Timisoara, Romania

[*] Department of Computer Science
Hofstra University, Hempstead, NY 11549

April 1 2008

# Cities of the Future: Employing Wireless Sensor Networks for Efficient Decision Making in Complex Environments

Alex Doboli, Daniel Curiac[†], Dan Pescaru[‡], Simona Doboli[*], Wendy Tang,
Costantin Volosencu[†], Michael Gilberti, Ovidiu Banias[†,] Codruta Istin[‡]

Department of Electrical and Computer Engineering
State University of New York at Stony Brook, Stony Brook, NY 11794-2350
Email: adoboli@ece.sunysb.edu

[†] Department of Automatics and Technical Informatics
"Politehnica" University Timisoara, Timisoara, Romania

[‡] Department of Computer Science
"Politehnica" University Timisoara, Timisoara, Romania

[*]Department of Computer Science
Hofstra University, Hempstead, NY 11549

*Abstract*

*Decision making in large scale urban environments is critical for many applications involving continuous distribution of resources and utilization of infrastructure, such as ambient lighting control and traffic management. Traditional decision making methods involve extensive human participation, are expensive, and inefficient and unreliable for hard-to-predict situations. Modern technology, including ubiquitous data collection though sensors, automated analysis and prognosis, and online optimization, offers new capabilities for developing flexible, autonomous, scalable, efficient, and predictable control methods. This paper presents a new decision making concept in which a hierarchy of semantically more abstract models are utilized to perform online scalable and predictable control. The lower semantic levels perform localized decisions based on sampled data from the environment, while the higher semantic levels provide more global, time invariant results based on aggregated data from the lower levels. There is a continuous feedback between the levels of the semantic hierarchy, in which the upper levels set performance guaranteeing constraints for the lower levels, while the lower levels indicate whether these constraints are feasible or not. Even though the semantic hierarchy is not tied to a particular set of description models, the paper illustrates a hierarchy used for traffic management applications and composed of Finite State Machines, Conditional Task Graphs, Markov Decision Processes, and functional graphs. The paper also summarizes some of the main research problems that must be addressed as part of the proposed concept.*

## 1. Introduction

From a functional point of view, cities are complex systems in which resources and infrastructure are continuously distributed to activities of the local economy and inhabitants [CACCIT, NAE, Dineen (2000), Dion and Yagar (1996), Fehin (2004), Felici *et al* (2006), Greene (2007), Hull *et al* (2006), Korkmaz *et al* (2004), Licalzi and O'Connell (2005), Murty *et al* (2007)]. The distributed resources are electricity, water, heating gas, parking space, etc., but also commodities like dining services, lodging, hospitals, entertainment facilities, and so on. The city infrastructure includes roads, sewage, water pipes, power grid, etc., in general, any means that enables the continuous transferring of resources to satisfy emerging needs. In this context, a city must have sufficient decision making capabilities for efficiently and continuously allocating resources through the infrastructure to satisfy demands. Efficiency is defined with respect to many metrics like monetary cost, quality of services, response time, safety, overhead (related losses), etc.

The quality of decision making represents a trade-off between the effectiveness and comprehensiveness of the decisions, and the related costs. Global decisions can be more efficient as they involve detailed analysis of large amounts of data. However, they are time and money-wise the most expensive as they require lengthy procedures executed by large personnel. In contrast, local decisions are much cheaper but they rely on small amounts of data, and use much simpler decision making procedures. In traditional urban set-ups, decision making remains arguably a lower efficiency process, which rarely relies on state-of-the-art technology for rapidly identifying the best options for a given situation. It has been recently suggested that modern technology, including ubiquitous data collection through sensors, automated analysis and prognosis, and online optimization can offer intriguing new capabilities for effective decision making in large urban environments while keeping the related costs low [Campbell *et al* (2006), Chakrabarty *et al* (2003), De *et al* (2005), Dubois *et al* (2003), Eisenman *et al* (2007), Greene (2007), Hull *et al* (2006), Lee *et al* (2006), Jain *et al* (2004), LeBrun *et al* (2005), Lee *et al* (2006), Murty *et al* (2007), Yoneki (2005)].

Popular applications like, large scale lighting control and traffic management, suggest that the main characteristics of decision making procedures include (i) flexibility, (ii) autonomy, (iii) scalability, (iv) efficiency, and (v) predictability. (i) *Flexibility* is the capability of a deployed system to address situations that are not predicted or addressed off-line. This is important due to the large variety of situation that occur in real-life. (ii) *Autonomy* refers to the capacity of a system to operate with minimum human intervention. This is critical as the cost of servicing decision making systems by humans might prohibit their deployment in large cities. Moreover, having humans tightly "embedded" in the decision making loop also increases the response time and can reduce the reliability due to inevitable human errors. (iii) *Scalability* characterizes the effectiveness of decision making methods with increasing amounts of demands, resources, and infrastructure. This is crucial for urban set-ups, which involve huge numbers of correlated activities, demands, resources, and infrastructure. (iv) For on-line and autonomous decision making systems, *efficiency* represents the quality of the decisions automatically selected during operation (online) as compared to the quality of offline decisions based on the entire state information being available. High efficiency is vital in real-life applications. (v) *Predictability* is the ability of devising automated procedures that estimate accurately and in real-time the quality of decision making alternatives. Predictability is a key enabling component for any decision making algorithm.

In addition to the five criteria, another requirement refers to the capability of making decisions that comprehensively tackle different applications that are correlated or might become correlated in certain conditions. For example, it is likely that the illumination intensity control along roads must be correlated to the traffic level on the illuminated roads. Hence, the control procedures for street lighting control have to interact with the decision making procedures for traffic management, such as traffic signal control. This interdependency can be analyzed off-line and incorporated into the system before operation provided that there is a common model and specification formalism for the two separate decision making applications. A more complex situation occurs if correlations dynamically emerge only in certain conditions, which are hard to predict a-priori. Then, the decision making system ought to recognize any newly created dependencies, and provide in real-time a strategy for their common resolution. Traditional decision making systems incorporate mostly only static procedures, and thus are likely to fail in unpredicted situations. For example, the control of the heating gas pipes is often uncorrelated to the traffic flow. However, if gas leakages occur then the lighting intensity should be diminished to reduce the risk of explosions, which further requires the adjustment of the traffic signals, so that the traffic flow through that zone decreases. The identification of all correlations that can occur, while conceptually possible, is in practice infeasible due to the many situations and conditions that are possible in large cities. Moreover, it unnecessarily complicates the control procedures considering that many of the possible situations might actually not occur in practice.

A number of recent research projects have focused on the topic of using state-of-the-art computing and communication technology, including wireless sensor networks, for various applications in large urban environments. The tackled applications include public transportation [Yoneki (2005)], traffic monitoring [Hull *et al* (2006), Lee *et al* (2006)], traffic light control [Cunningham *et al*], social networking and VoIP [Murty *et al* (2007)], data gathering [Campbell *et al* (2006)], communication [Kansal *et al* (2004)], just to mention a few. The present research focuses on efficient and comprehensive data acquisition, wireless routing protocols, programming paradigms, embedded hardware, and security. An intriguing opportunity is to device mechanisms that allow integrating and co-optimizing the correlated applications rather than the more traditional method of tackling them independently. This not only expands the optimization space, but also increases the reliability of decision making in new situations. Moreover, it is important to have a decision making approach that is simultaneously flexible, autonomous, scalable, efficient, and predictable.

This is challenging as the performance of flexible and autonomous systems is hard to predict in the general case. Moreover, large systems are difficult to optimize through on-line decision making procedures.

This paper proposes a new decision making paradigm for optimizing the continuous, real-time allocation of resources to satisfy demands in large urban environments. While the proposed paradigm is general, the paper refers to street lighting and traffic control as two illustrating case studies. The paradigm is structured as a *semantic hierarchy* in which different decision making models and strategies coexist and interact to produce flexible, autonomous, scalable, efficient, and predictable decisions. The hierarchal structure is scalable as only a reasonably large number of modules *collaborate* at each semantic level for reaching global decisions. Flexibility and autonomy is achieved by having reactive models at the bottom of the decision making hierarchy. Hence, unexpected sequences of events can be accommodated during operation. Efficiency and predictability results from having the reactive behavior constrained by the upper semantic levels, which use more deterministic models, such as data flow graphs, and Markov Decision Processes [Feinberg (1994), Feinberg (2002)]. The upper levels compute the limits within which the lower, reactive decision making modules must operate, so that the overall goals and constraints of the application are not violated. The lower semantic levels are in constant interaction with the environment and acquire data about real situations. This data is aggregated and propagated to the upper semantic levels, where it is employed for global decision making. Thus, the paradigm incorporates feedback mechanisms between the semantic levels, so that the upper levels constrain the lower levels while the latter provide feedback on the effectiveness of the constraints. Finally, the paper also introduces the defining elements of a specification formalism for integrating different but related applications. Various kinds of interactions between decision modules, such as collaboration, competition, guidance, and enabling interactions, are defined based on the goals, capabilities, inputs, and outputs of each module. This represents a uniform description style for different applications.

While the proposed decision making paradigm is applicable to other theoretical formalisms too, this paper refers to a semantic hierarchy based on Finite State Machines (FSMs), Conditional Task Graphs (CTGs), Continuous Time Markov Decision Processes (CTMDPs), and Functional Graphs (FGs). The hierarchy is exemplified for coordinated traffic signal control, a main application in modern cities [Gartner (1983), Gartner *et al* (1992), Wahle and Schreckenberg (2001)]. It implements *quantitative* decision making in which decisions at successively higher semantic levels are used to cover increasingly broader geographical areas: (i) FSMs implement reactive control guided by signals coming from traffic sensors and cameras as well as neighboring FSMs, (ii) CTGs schedule the related activities over time for different traffic conditions, thus conduct optimization of signal sequences, (iii) CTMDP use macroscopic descriptions of the system to conduct scenarios-specific optimizations (hence, time is abstracted away from decision making), and (iv) FGs regulate the global allocation of resources and infrastructure to demands, hence taking out the impact of scenarios. The generality of the paradigm is motivated by summarizing a *qualitative* decision making process, in which the semantic hierarchy uses FSMs, fuzzy logic, and expert systems for decision making.

The paper has the following structure. Section 2 introduces two applications (city lighting control and traffic management) to illustrate the importance of infrastructure management in cities. Section 3 defines the main characteristics of decision making in large urban environments, and Section 4 presents the proposed decision making approach. Section 5 introduces the specification model for integrated decision making, and Section 6 enumerates some of the main research issues related to the proposed concept. Finally, conclusions are offered.

## 2. Applications

A myriad of applications can use the more efficient decision making capabilities that are enabled through ubiquitous data collection, automated prognosis, and intelligent advising. Applications differ depending on the geographical localization of decision making, and the kind of human interaction involved in the process. The first criterion defines not only the "globality" of the resulting decisions, but also it quantizes the hardness of the process, as selecting optimal decisions for smaller, geographically localized areas is arguably simpler than that for large, distributed areas. The second criterion centers on the human perspective to distinguish situations where humans are only passive participants to the process, and situations where humans participate to the decision making loop. In the first case, decision making actually results in an actuation activity, which produces a change in the way resources and infrastructure is used. In the latter case, the activity results in additional knowledge, which humans might use for taking more educated decisions. Two applications are discussed next: ambient lighting control and traffic management.

## A. Ambient lighting control

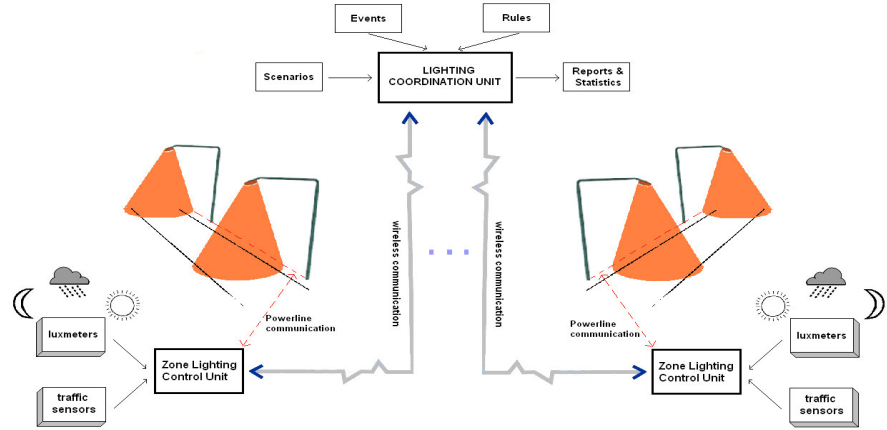

*Figure 1*: Ambient lighting control system

Ambient lighting control applications must automatically adjust the lighting intensity in public spaces so that optimal illumination is offered while minimizing the energy consumption and maximizing reliable operation. These systems must operate with no or little human intervention. They automatically turn on, off, and adjust the illumination intensity in public areas, streets, buildings, parks, and so on. The illumination intensity can be adjusted depending on the time of the day (e.g., dawn, dusk, etc.), weather conditions (sunny, cloudy, full moon), presence of obstacles, number of pedestrians, the nature of the activities carried on (e.g., walking, bicycle riding, etc.), and more. Other capabilities could be related to minimizing energy costs and improving safety, such as by switching to different energy providers during power outages, or selecting the providers that offer the cheapest energy. Similar applications include automated monitoring of the air quality and measuring the noise levels [Santini and Vitaletti (2007)]. Such applications not only reduce living cost and save energy, but also offer better quality-of-life in a city.

The next paragraphs offer more insight into an *ambient lighting control system* (ALCS) that attempts to offer optimal illumination for a given energy budget. Figure 1 presents the ALCS structure. Optimal illumination is defined as a certain illumination intensity at the street level. Each lamp post is equipped with local sensors (e.g., luxmeters, motion detectors, orientation sensors, distance sensors, and other traffic related sensors), and a local controller to decide the illumination intensity. All local controllers of the lamp posts in a small zone (e.g., a street) are connected into a network to coordinate their decisions. Moreover, zones are linked together through wireless connections, and *Lighting Coordination Unit* (LCU) coordinates the different zones, if more global decisions ought to be made, such as to illuminate during events, holidays, etc. Moreover, if the energy cost budget is fixed, LCU divides the budget to the related sub-areas, so that the overall consumption stays within the budget. Sub-areas can renegotiate their assigned budget. LCUs also collect data for reports and statistics, and identify behavior patterns for different situations.

To provide scalability and flexibility, ALCS implements a two-level control hierarchy, as in Figure 1:
- *Local controller*: The controller decides the illumination intensity of a lamp depending on the conditions sensed through the sensors, and the interactions with the controllers of the neighboring lamp posts. For example, if one of the lamps ceases to operate, the illumination intensity of its neighbors is increased, so that they can cover the entire area. Similarly, lamps might be turned off to save energy, if the illumination level exceeds the needed values.

Decision making implements a multi-mode control with specific control laws operating for each of the functioning modes. In a simpler case, decision making represents a reactive behavior in which the exceeding of pre-set threshold values (e.g., a neighboring lamp going off) results in switching to a new state in the decision making controller. For each of the states, the inputs from the sensors are used to control a simple "bang-bang" control algorithm, in which the lighting intensity increases if the sensed value is less than the optimal value, and decreases otherwise. Other control algorithms like PID control are also attractive.

A complementary approach to local control implementation is using a fuzzy logic [Driankov *et al* (1996), Farinwata *et al* (2000), Passino and Yurkowich (1998)]. The fuzzy logic controller assesses qualitatively the inputs coming from sensors and neighboring controllers, and then selects the actuation outputs using the inputs in a set of predefined rules (which define the fuzzy logic controller). For example, if all inputs are large then a specific response is produced. Similar reactions correspond to all inputs being small, and so on for all qualitative combinations of input values. This approach can handle well the imprecision of the sensors outputs, as well as cases when the lighting intensity must be adjusted for hard-to-predict values, like the traffic volume.

- *Lighting Coordination Unit (LCU)*: The upper-level controller takes more long term decisions that relate to broader areas. The traditional approach implements a centralized controller for LCU. The decision making algorithm uses all requests from the zones to compute the energy resources allocated to each zone using optimization methods, like Linear Programming or heuristic techniques [Reeves *et al* (1993)]. While well understood, centralized control does not scale well for large systems as a large overhead is involved. Instead, decentralized control involves only a small number of neighboring zones, which decide resource allocation based on greedy minimization (maximization) of priorities [Sinopoli *et al* (2003)].

    LCU also maintains connection with all related decision making systems within the city limit (like traffic management, disaster management, etc.), to responds efficiently to unscheduled events, e.g., traffic jams, high levels of environmental pollution, power breakdowns, and so on. More details on connecting the different decision making systems are given in Section 5.

## B. Traffic management

Improving the efficiency of traffic is another important application of intelligent decision making in urban environments [Cunningham *et al*, Yoneki (2005), CALCCIT]. Poor traffic management results in long term waiting times, increased gas consumption, and pollution. Efficiency parameters might include the number of vehicles serviced in a given time interval, the time required to cover a certain trajectory, the pollution and noise levels due to traffic, the maximum number of vehicles that can be serviced without vehicle queuing as well as parameters that define special situations such as offering the shortest travel time to police cars, ambulances, fire-trucks, etc. Tasks related to traffic management include finding the optimal route for vehicles, identifying and warning about traffic "hot spots", re-routing in case of accidents and other unforeseen obstacles, redirecting traffic to avoid high pollution levels in a zone, and more.

Decision making for traffic management must optimize the related goals (e.g., efficient traffic, reduced gas consumption, minimal noise levels, etc.) through either centralized or localized strategies [Cunningham *et al*]. Centralized algorithms express global goals and evaluate decision on a central processing unit, which also collects the data sampled by the individual sensing units. Centralized control, the traditional approach, offers good results for predictable traffic flows, however, its main limitations include low scalability and poor reliability. Localized algorithms rely on a small number of neighboring units for taking decisions, which only affect a small area. In addition, they include a mechanism for percolating the local interactions to the global area, so that large scale decision making is also possible. This approach is potentially more flexible in handling less predictable traffic situations, scalable, and reliable, nevertheless, its main challenge is the difficulty to correctly estimate global characteristics about functioning and performance.

Similar to the previous example, the decision making approach proposed for traffic management is hierarchical, in which the lower control levels are localized methods and the upper levels are more global.

The local controllers handle the street-level traffic in contrast to the global-level procedures, which identify strategies at the area level to guarantee the goals and performance constraints of the application. Compared to street lighting control, the decision making hierarchy is in this case higher as traffic management involves a wider geographical scale and operates in more diverse situations.

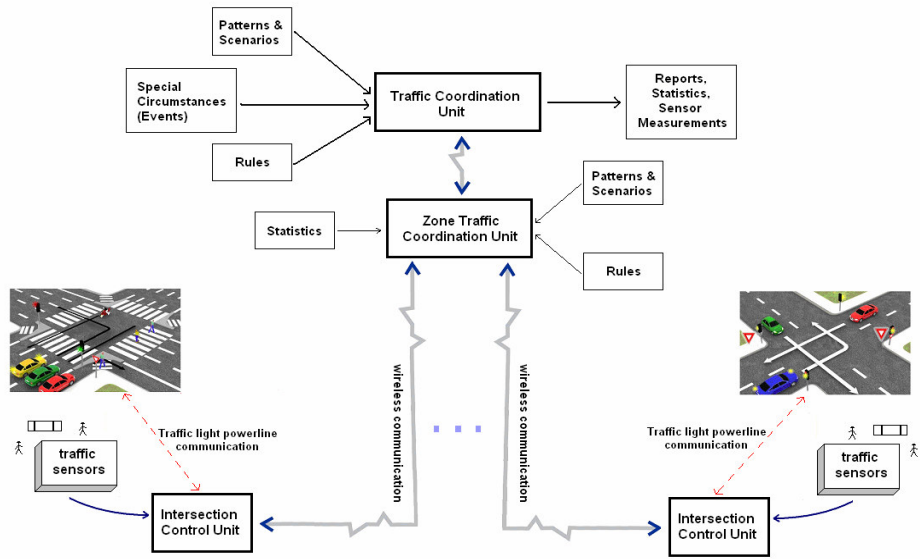

*Figure 2*: Traffic management decison making system
(Acknowledgement: The two sub-figures representing traffic situations are due to http://www.permisulauto.ro)

Figure 2 depicts a three-level hierarchical system, even though more decision making levels can be added:
- *Intersection Traffic Unit* (ITU): ITU is the controller of the traffic light that supervises each street intersection. ITU receives inputs from sensors, such as inductive loops embedded in the pavement, video cameras, etc., on the traffic intensity in each direction, the length of the vehicle queues, the time taken for a vehicle to move through the intersection, and more [Cunningham *et al*]. The ITU behavior can be extended with actions for preferred "vehicles", like bicycles or pedestrians, which ought to have higher priority to pass through the intersection. ITU might also include a "default" behavior which is executed whenever the controller fails to converge on a predictable operation.
- *Zone Traffic Coordination Unit* (ZTCU): ZTCU coordinates the operation of ITUs in neighboring intersections to optimize the traffic through each of them by coordinating the decisions of the individual ITUs, redirecting the traffic flow, etc. ZTCU receives data from the participating ITUs, such as traffic characteristics, vehicle queues, moving obstacles, and so on. It might also receive inputs from the upper most Traffic Coordination Unit (see next item) about the targeted traffic efficiency goal, like avoiding long queues and minimizing the delay of vehicles. ZTCU computes optimal strategies for different traffic situations (patterns) in the intersections.
- *Traffic coordination unit* (TCU): TCU operates at the global level by setting the parameters of the ZTCUs. Deterministic policies are used for decision making at this level. TCU also adapts the traffic flow to incorporate out-of-the-order, global situations, such as during the passing of a rapid intervention unit or fire truck that must traverse quickly through a large area.

Sections 3 and 4 give more details on the methods proposed for each decision making level.

## 3. Proposed Decision Making Concept

Decision making in complex urban environments must offer the following main capabilities:
- *Flexibility*: During operation, the system must be flexible enough to accommodate new situations, objectives, and requirements, which are hard to predict off-line [Saleh *et al* (2002)].

- *Scalability*: Scalability is the property of the control algorithms to perform well for increasing numbers of service requests, resources, and infrastructure. There are many reports that indicate that decentralized methods, which use only information localized in time and space, handle better complex systems than centralized methods [Sinopoli *et al* (2003)].
- *Autonomy*: Decision making should minimize human intervention. This is important not only for the scalability requirement of decision making but also because of the higher costs and lesser, if human intervention is involved.
- *Efficiency*: To be viable in real-life, decision making must be effective with respect to metrics, like short service time, high service quality, low costs, good reliability and safety, maximum number of satisfied services, and many more. Moreover, for flexible and autonomous decision making, efficiency also captures that quality of adapted, online decision making as compared to offline decision methods in situations when all state information being available.
- *Predictable*: For autonomous and efficient operation, the characteristics of decision making ought to be predictable. For example, the resulting service time and costs of local decision making strategies should be predictable at a global level and over time. Also, control techniques should not lead to deadlock, thus the set goals should be reachable by the local strategy.

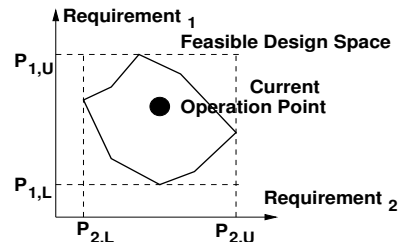

*Figure 3*: Definition of flexibility

The five properties can be measured using the following metrics inspired by the similar metrics used in manufacturing engineering and parallel systems:
- *Flexibility*: In [Kazmer *et al* (2003)], flexibility (in engineering design) is defined as the ratio of the feasible design space over the range of the specification. A similar definition can be used for decision making in urban environments. Figure 3 illustrates the definition for two performance requirements. The figure assumes that a specification range $[P_{i,L}, P_{i,U}]$ is defined for the requirements $P_1$ and $P_2$. Moreover, some of the requirement values cannot be realized in practice, thus, a space of feasible design points is defined around the current functioning point. A larger feasible design space determines a higher flexibility of the system. The definition can be easily extended for more requirements.
- *Scalability*: According to [Jogalekar and Woodside (2000)], the scalability of a distributed systems with respect to attribute $P$ reflects the amount of extra resources that ought to be spent for improving P's value from $P_1$ to $P_2$. If we denote $Cost_1$ the resource cost for achieving performance $P_1$, and $Cost_2$ the resource cost for having performance $P_2$, then scalability is the following ratio:

$$\text{Scalability}(P_1, P_2) = \frac{P_1 \cdot \text{Cost}_2}{P_2 \cdot \text{Cost}_1}$$

- *Autonomy*: Autonomy depends on the amount and quality of human input required to assure a system's correct functioning under a broad range of changing requirements, which act over a specified geographical area and over an interval of time. The following formula presents a qualitative expression of autonomy:

$$\text{Autonomy} = \int_{T_i}^{T_f} \int_{V_i}^{V_f} \int_{PU}^{PL} (\text{Human effort}) \, d\text{Performance} \, d\text{Area} \, d\text{Time}$$

where performance **P** changes from $P_U$ to $P_L$, the geographical area border is defined by $V_i$ and $V_f$, and the time interval between moments is $T_i$ and $T_f$.

- *Efficiency*: Flexibility and autonomy are obtained at the expense of extra resources and infrastructure being spent in contrast to the optimal, offline solution. For example, in dusk conditions, an adaptive lighting system might be less efficient than a system specifically designed for such conditions. Nevertheless, it is not feasible to have a dedicated optimal design for each of the conditions that might appear in real-life. For performance **P** and range $[P_L, P_H]$, efficiency is the average overhead spent by the adaptive system to achieve each of the performance values in the range:

$$\text{Efficiency}(P_L, P_H) = \frac{\int_{P_L}^{P_H} (P_{addaptive} - P_{single-value}) dP}{P_H - P_L}$$

- *Predictability*: Developing flexible and autonomous decision making strategies depends on having techniques that can estimate correctly the feasibility and quality of alternative decisions. Thus, the error between the estimated and the real values must stay within a tolerable limit.

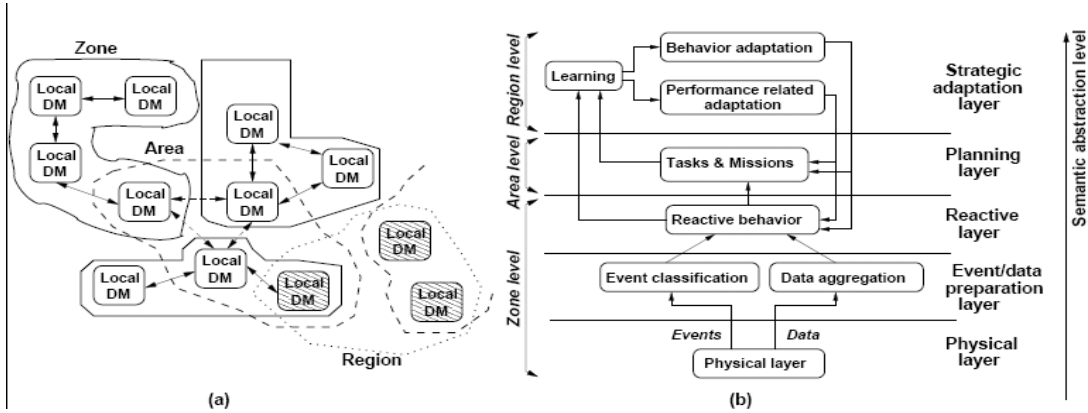

*Figure 4*: (a) The wide-scale decision making model and (b) the local control approach

The proposed decision making model attempts to address the five main capabilities enumerated before. Figure 4 illustrates the concept. The approach realizes control through localized data acquisition and collaboration between the interrelated sub-systems. This is critical for having a scalable decision process. Depending on the goals, systems organize themselves into clusters. Each cluster might decide to elect virtual leaders used in collaborating with other clusters. Figure 4(a) shows decision making for three successively larger geographical parts: three local zones (shown in solid line), two larger areas (in dashed lines) and one broad region (in dotted line).

Clusters correspond to increasingly higher semantic abstraction levels, each level using different formal representation models. For example, the lowest levels operate in terms of events and basic signals/data and the actions corresponding to them. This is typical to reactive behavior. The intermediate control levels reason in terms of optimizing the execution of multiple tasks over time but without analyzing the physical events and data. Finally, the upper levels consider global and qualitative decision making which are optimal across long period of times (hence, the temporal issues are eliminated from the model). Hence, the control algorithms for clusters covering broad regions remains scalable due to the more abstract information being used during analysis and decision making.

Figure 4(b) illustrates the hierarchy of semantic abstraction levels. Each local decision making module has a layered structure to offer a gentle transition from reactive behavior to fully predictable functioning.

- The *physical layer* is mostly concerned with the acquisition of physical signals (such as voltages coming from sensors) and the conversion of these signals into data and events. The physical layer utilizes reconfigurable hardware (e.g., reconfigurable analog and digital circuits) to optimize the efficiency of data acquisition and conversion. For example, faster hardware circuits are used if the speed requirements of data acquisition increase due to faster processes that have to be tracked, or more precise signals processing circuits are needed if the accuracy needs to be increased.
- The *events/data preparation layer* aggregates data and classifies events. Multiple sensors might sense the same data or fragments of the same data. The redundant data can be used to improve the precision of sensing by eliminating noise. Also, comprehensive knowledge is assembled together from the data pieces sampled by each sensor. Event classification is important in situations where different event types are handled by different processing algorithms, like specific responses to different vehicle types (e.g., trucks, bicycles, passenger cars, etc.).
- The *reactive layer* produces responses to the input data and events.
- The *planning layer* controls the execution of tasks, so that the set goals are maximized and also the using of the infrastructure and resources is optimized. The planning layer guarantees that the constraints set for the system operation are met, including the deadlines and energy budgets of the individual tasks. Therefore, the behavior and performance of the planning layer must be fully predictable in contrast to the reactive layer, which is flexible but hard to estimate.
- The *strategic adaptation layer* identifies repeatable patterns which can be used to optimize the functioning of the system with respect to goals, resources, and infrastructure. The identified patterns include sequences of input data, correlations between input data, correlations of events, optimal control sequences for certain conditions, and so on.

Section 4 offers more insight into the decision making components.

## 4. Efficient Decision Making

Efficient and robust behavior in hard-to-predict environments does not only imply that the autonomous system can react to unexpected events and conditions but also is capable to meet all performance requirements, including real-time constraints, bandwidth limitations, energy constraints, speed requirements, precision needs, safety, and so on. As we explained in the introductory sections, this is a hard problem, and there are no predictable control algorithms for integrated mixed-domain reactive behavior.

We propose a predictable yet flexible decision making concept based on (i) a *semantic modeling hierarchy* to represent the performance needs and capabilities of a decision strategy at different levels of abstraction, and (ii) *a joint top-down and bottom-up constraint transformation mechanism* along the semantic hierarchy for guiding the dynamic reconfiguration process and signaling when a set of requirements is insatisfiable for the current set of resources. If the application requirements cannot be met then the autonomous decision making system can decide to shift some of the load to other systems (if they are available), or to trade-off the response quality, such as the response time and accuracy. Tackling the load shift problem without using a centralized control algorithm (which would pose serious scalability limitations) is beyond the scope of this work, but we plan to consider modern techniques for decentralized and local control when researching integrated dynamic reconfiguration, such as the algorithms in [Reynolds (1987), Brogan (1997), Ulam (2007), Brock (1999), Pereira (2003), Hsieh (2007), Sigurd (2006)].

The main features of the two components of the proposed predictive control mechanism are as follows:
- *Semantic modeling hierarchy*: The semantic hierarchy is an abstraction model, which offers a smooth transition from the fully reactive behavior at the node level and the deterministic operation at the application level. The hierarchy includes several abstraction levels, so that each higher level is less flexible in tackling unknown conditions but is more predictable performance-wise. Figure 5 shows a semantic hierarchy with five layers, in which the two bottom layers represent reactive behavior, and the top-most layer offers a performance predictive description of the system as a task graph with data dependencies.
- *Top-down and bottom-up constraint transformation*: A set of top-down and bottom-up constraints enforce the consistency between consecutive layers. The top-down constraints are defined by an upper level for the level immediately below, so that the behavior at the lower level remains within the "bounds" of the behavior at the upper level. The bottom-up constraints are defined from a lower level to the next upper level, and express the amount of performance "violation" (performance needs) that must be considered when re-computing the resource management and control algorithm at the next upper level. The adjustment (propagation) of top-down and bottom-up constraints is performed continuously at runtime.

Hierarchical control of robots, a similar though less complex problem, is proposed in recent work. For example, a two level control hierarchy is proposed in [Pereira (2003)], where the lower level is a finite state machine guided by events and the upper level is a decentralized control method between an autonomous mobile system and its immediate neighbors. Local control is based on the potential field method. Similar approaches are discussed in [Ulam (2007), Ulam (2007b), Hsieh (2007)]. The proposed concept differs in several main issues from this work. First, the semantic modeling hierarchy includes more levels, including layers which use stochastic models, conditional task graphs, and task graphs with data dependencies. Second, the proposed hierarchy also aims at expressing end-to-end dynamic reconfiguration of the design and not only reactive functionality, like in the existing approaches. Hence, the proposed models will tackle a wider set of performance requirements, in contrast to existing approaches which mostly consider only convergence to the application goal and sporadically also timing constraints. Third, the predictive mechanism is based on a runtime propagation of top-down and bottom-up constraints and the re-computing of the corresponding control decisions. There is no such adaptation mechanism in the other methods. Existing techniques perform only an off-line static performance analysis, mostly related to the convergence and stability of the used control methods.

The two components are discussed in the following two subsections.

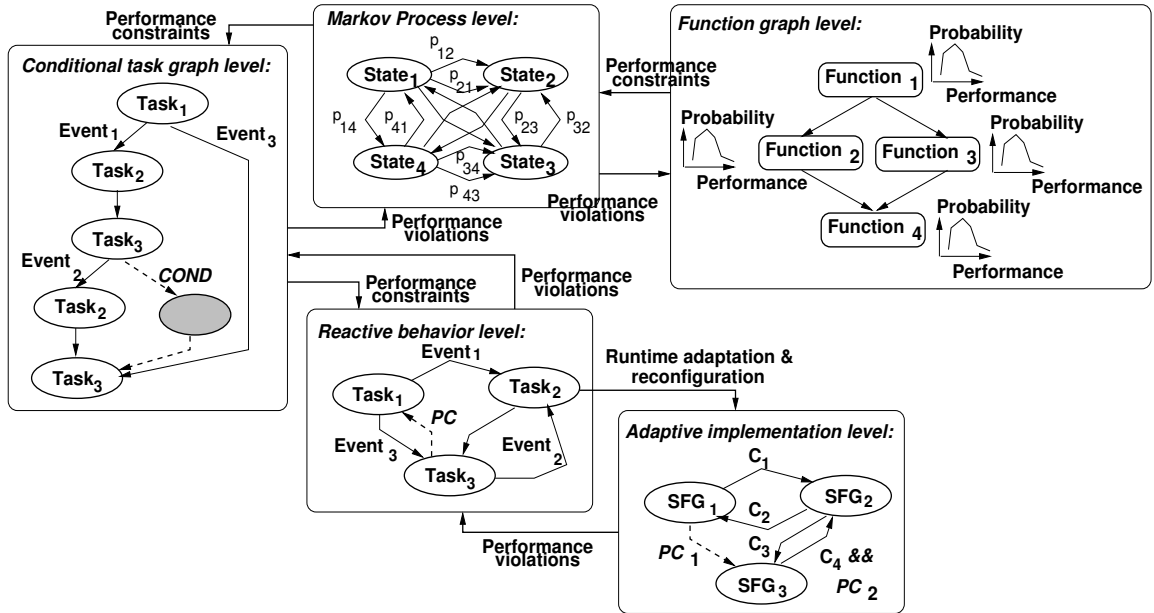

*Figure 5*: Semantic modeling hierarchy for large-scale decision making

**4.1 Modeling Hierarchy for Predictive Operation**

Figure 5 illustrates the suggested semantic modeling hierarchy for predictive decision making. The two bottom levels describe the reactive operation of the decision making system, and the three upper levels provide a smooth transition towards predictable control. The top most level defines the functionality as a task graph with a predictable and evaluatable performance. The five model layers are as follows:
- *Adaptive implementation level*: This layer expresses the behavior of the dynamically modifiable system configuration. This layer defines the reactive adaptation of the implementation to dynamic conditions. Each node is a signal flow graph (SFG) defining a reconfigurable design (analog, hardware, and software). The transitions between SFGs are controlled by external events and other conditions set during operation. This level corresponds to the physical and events/data preparation layer in Figure 4(b).

- *Reactive behavior level*: This layer defines the reactive functionality of a system as FSMs. Each node is a functional task (e.g., trajectory computing, data communication, sensing, etc.). State transitions are controlled by events that are set during operation. Each task is represented as an SFG graph at the adaptive implementation level.
- *Conditional task graph level*: Conditional Task Graphs (CTGs) define various execution traces for the FSM at the reactive behavior level. A certain path of a CTG is executed depending on the runtime values of the conditions (events) labeling the graph arcs. The graph nodes are either individual tasks in the lower level FSM, or FSM portions, which are "lumped" together as a single node to reduce the size of the CTG. The conditions describing different paths are either single events in the FSM, or logic expressions based on the events.
- *Markov Process level*: For each task (function), its execution traces (which is a complete path through the CTG) define a Markov Decision Process. This is suggested by experimental observation that the similar behavior of many computing systems can be approximated as a Markovian Process [Benini (2000), Qiu (2001), Kallakuri and Doboli (2007)]. Each trace is a state. Transition rates between states represent the probability of executing that trace in the CTG.
- *Function graph level*: The application is described at the top-most level as a function graph in which a node is a specific function (task), and arcs define the order of performing the tasks. Nodes are unconnected, if no specific order is required. The performance of each task (e.g., its execution time) is described as a distribution that corresponds to the underlying Markov Decision Process. Please note that the system behavior is predictable at this level because the model can be deterministically evaluated.

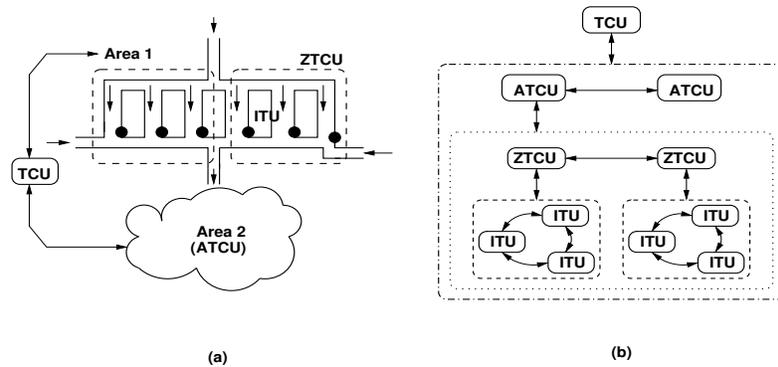

*Figure 6*: (a) Simple case study and (b) hierarchical decision making structure

The hierarchical decision making in Figure 5 operates as follows. The two bottom most levels employ event driven decision making mechanisms, such as Finite State Machines. The *reactive behavior* level description switches from one task to another depending on the occurrence of events. For example, the system switches from `Task`$_1$ to `Task`$_2$, if `Event`$_1$ occurs during operation. Each controlled unit executes its own controller. The *adaptive implementation* level refers to the capability of adapting online the implementation of a certain task, such as switching from a reconfigurable implementation to another under conditions `C`$_i$, which relate to implementation parameters, like the available energy, memory, communication bandwidth, etc. of a device. Several individual controllers might decide to collaborate by building collectively a shared description at the *conditional task graph* (CTG) level. The description defines how the individual controllers use jointly any shared resources to achieve common objectives, such as the access over time (schedule) of vehicles accessing intersections. The CTG description includes decisions specific to different conditions, such as various traffic loads. Please note that CTG decision making is at the level of small local areas rather than individual controllers. Next, the CTG descriptions of the correlated areas are used to build description covering broader areas, such as those at the *Markov Decision Process* level. Finally, the last level is at the *function graph* level, and is based on information collected from Markov Process level descriptions for several interrelated areas. This step distributes the overall goals into goals for the individual sub-systems using information from the Markov Decision Process level.

The remaining part of the subsection presents the top-down bottom-up constraint transformation for predictive operation. The five consecutive layers in Figure 5 are related with each other through top-down and bottom-up constraints. The top-down constraints reflect the performance requirements that are imposed by the application, and which must be satisfied during operation. The bottom-up constraints (called performance violations in the figure) capture the amount of change in the control algorithms, if the current reactive implementation (at the bottom two levels) cannot operate in unpredictable conditions while the application requirements are satisfied.

We will exemplify the concept of top-down and bottom-up constraints by looking at the bottom two levels in Figure 5. Depending on the performance requirements expressed at the reactive behavior level (e.g., the time constraint for task execution, precision, etc.), the control algorithm switches among the SFGs of the adaptive implementation level. If the constraint is not met then the control algorithm switches to the SFG for a more efficient implementation, and vice versa, if the required performance is exceeded. Similar control strategies have been recently proposed in [Lu (2001), Zhang (2002)a], but the methods are for software only and cannot handle mixed-domain constraints. If a performance constraint is violated then the control algorithm might decide to transition immediately to a safe implementation (as shown by the dashed line arc labeled as $PC_1$ in the figure). For example, if the system is running out of time then it might decide to immediately switch to the most efficient state without going through any intermediate states.

Similarly, the lower level will provide bottom-up feedback to adjust the transitions of the FSM at the reactive behavior level, in case the imposed performance constraints are violated. Such a "short cut" transition is indicated by the dashed line arc labeled as `PC` in the figure. Short cut transitions are also introduced by the performance constraints originated from the conditional task graph level.

At the conditional task graph level, new nodes and edges are inserted into the graph for providing a predictive behavior. In the figure, the nodes are shaded and arcs are shown with dashed line. If performance constraints are not met then the resource management module might decide to perform a shaded node instead of the node corresponding to the functionality. The missed functionality might be allocated to another node, if there is a node available, or might be entirely skipped, if it does not violate the application requirements.

The remaining of the section illustrates two possible instances of the proposed semantic decision making hierarchy: (A) traffic management based on FSM – Conditional Task Graphs – Markov Processes – Function Graph hierarchy, and (B) light control using Fuzzy Logic – Knowledge Based Rules. While the first hierarchy relies more on quantitative decisions, the second semantic hierarchy is more geared towards knowledge based (qualitative) decisions.

**4.2 Traffic Management using FSM-CTGs-Markov Descision Processes-Function Graphs hierarchy**

Figure 6 shows the simple traffic management case study that has been explored and the corresponding semantic hierarchy. Figure 6(a) presents a street network, in which arcs indicate the flow of traffic. The goal of decision making is to maximize the traffic flow that passes in a given time `T` through the area. Traffic lights (shown as black bubbles) are positioned at each intersection, and each is controlled by its local ITU (Intersection Traffic Unit, see Section 2). Neighboring ITUs form a ZTCU (Zone Traffic Coordination Unit, see Section 2) that coordinates their operation. Moreover, ZTCUs of an area are interconnected and coordinated by an ATCU (Area Traffic Control Unit). Finally, ATCU are supervised by TCU (Traffic Coordination Unit). Figure 6(b) illustrates the resulting semantic hierarchy. ITUs, ZTCUs, and ATCUs can be statically defined, or can emerge dynamically during operation based on the actual traffic patterns and the best way of coordinating the traffic lights. We have assumed static definitions in this case study.

Figure 7 illustrates the two lower levels of the semantic hierarchy: the FSM level and the Conditional Task Graph (CTG) level. The figure also shows the interaction between the two levels.
- Each of the ITUs in the figure implements a FSM with three states (red, yellow, and green). Assuming cyclic traffic signals, arcs indicate state transitions which are taken after the FSM spends $\Delta T_{i,j}$ time units in the current state (the time is called *split* time). The time intervals are specific to each ITU and each transition, and are fixed by the ZTCU for optimizing the traffic flow. The schedule of the traffic signal repeats over a period $T$ (called *cycle time*). Neighboring ITUs communicate with each other to decide the delay between the monitored intersections (the delay is called *offset*).

- The ZTCU behavior is expressed as a Conditional Task Graph (CTG), which indicates the activities performed by two cars arriving at $ITU_1$ and $ITU_2$ in the figure. Activities include passing through the intersection, moving to the next intersection, and so on. Each of the activities is characterized by an execution time $T^{ex}_k$, which is estimated based on the data collected locally from the sensors at the traffic lights. ZTCUs set the cycle and split times of each intersection, and also find the initial delay values. Execution times differ for different traffic scenarios, e.g., light traffic (indicated as branch $L$) and heavy traffic (shown as branch $T$). The semantics of the application states that one and only one of the branches $L_i$ and $H_i$ is executed for a traversal of the graph. This semantics defines a conditional (multimode) behavior.

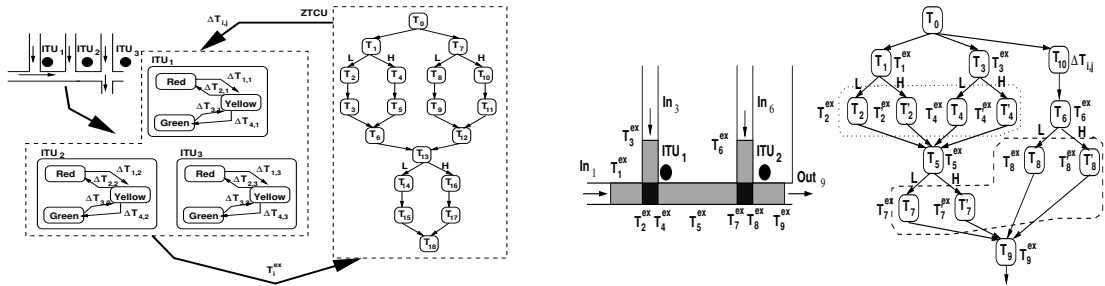

*Figure 7*: FSM – Conditional Task Graph hierarchy

Figure 7(b) presents the concrete CTG for a zone with two consecutive intersections controlled by the traffic signals $ITU_1$ and $ITU_2$. The graph presents the activities occurring during one cycle of the entire area, even though the cycle might include several cycles of the individual traffic signals. The process is repetitive after each area cycle. In this example, we assumed single-lane roads, which means that two cars coming through the intersections $ITU_1$ or $ITU_2$ cannot move simultaneously. Hence, certain road sections (shown as dark areas in the figure) act as shared resources for two or more cars, while other roads are dedicated resources. The model defines tasks $T_1$ and $T_2$ as being the travel of a vehicle entering the area from the left, tasks $T_3$ and $T_4$ as representing the moving of a vehicle coming from the top-left road, tasks $T_6$ and $T_7$ represent a car moving from top right, and so on. The resulting CTG is illustrated in the figure. Each task has two attributes: $N_i$ is the number of vehicles that traverses the section during one cycle, and $T_i^{ex}$ is the time required for the vehicles to pass through the section. Obviously, the two attributes are related to each other. In the graph, the tasks sharing the same resources (the road sections are depicted in dark) are marked with dashed and dotted lines, respectively. In addition, the dummy task $\Delta T_{12}$ represents the offset time between the two traffic signals.

The ZTCU for the area computes online the optimal scheduling of the tasks $T_i$ while tackling requirements such as, (i) to maximize the total number of vehicles passing though the zone, (ii) to minimize the time taken to cars to move through the zone, (iii) to minimize the fuel consumption in a certain area, (iv) to minimize the amount of polluting gases that are generated, and so on. To avoid continuous recomputing of schedules and to improve the flexibility of the approach, the ZTCU scheduler calculates traffic signal schedules for different traffic conditions, which determine the "behavior" (and thus the mathematical models) for the task attributes, such as parameters $N_i$ and $T_i^{ex}$.

Please note that each computed schedule corresponds to traffic conditions described by conditional behavior and the qualitative parameters $L_i$ and $H_i$ in the figure. For example, if the traffic load is low on the segment $1$ then the branch labeled as $L$ is taken after the task $T_1$ is completed, hence task $T_2$ is performed. If the traffic load is high then the branch labeled as $H$ is selected, and the task $T_2'$ is pursued. Tasks $T_2$ and $T_2'$ can never be performed simultaneously as they correspond to mutually excluding conditions (e.g., the traffic intensity cannot simultaneously be low and high). This generates eight different situations for the case study in Figure 7(b), and eight different optimal scheduling situations for the zone, such as the schedule $S_{(L,L,L)}$ with the total time $T^{tot}_{(L,L,L)}$ for the situation in which the traffic density is light through all three intersections. The actual conditions, which are labeled as **L** and **H**, can be different for different traffic lights, and can involve more than two situations. The concrete task scheduling for CTGs can be computed with algorithms similar to the one in [Eles *et al* (2000)].

Figure 8(a) illustrates the interaction between the local ITUs at each traffic signal and the CTG scheduling at the ZTCU for the interacting ITUs. The scheduling table computed online by the ZTCU is shown in the figure. The table includes the scheduling solutions for different scenarios that might arise in traffic. Please note that the actual semantics of the conditions $L_i$ and $H_i$ (e.g., what traffic conditions actually define them) are refined during execution depending on the variation of the parameters $N_i$ and $T_i^{ex}$. More conditions allow more precise tuning of the schedules, however, the computational complexity of calculating the optimized schedules is also higher [Eles *et al* (2000)]. The table presents the schedules for two situations: one in which traffic is light along all directions, and the second in which the traffic along direction two is more intense while the others remain low. The traffic schedules are computed so that the total number of cars passing through the zone is maximized while the timing along different directions is kept within fixed deadlines (e.g., to avoid large delays along certain directions due to enabling heavy traffic along other directions). The boxes in gray indicate some of the idle times along a direction.

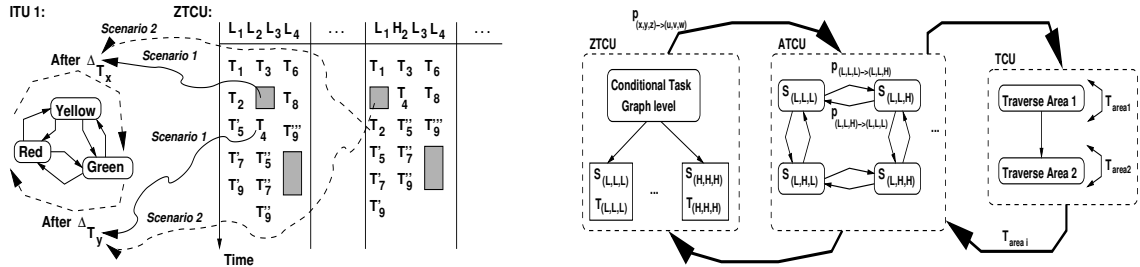

*Figure 8*: FSM - Conditional Task Graph – Markov Process – Functional Graph hierarchy

The information in the CTG scheduling table is mapped down onto the FSMs of each individual traffic signal. For example, for Scenario **1** (in which traffic is low along all directions), the time required for performing task $T_3$ plus the following idle time generate the timing constraint $\Delta T_x$, which states that after that amount of time the FSM must be in state **Green**, thus allowing traffic to progress along direction two. Similarly, for Scenario **2**, the time for task $T_1$ plus the following idle time define the constraint $\Delta T_y$, which states that the traffic signal ought to be in state **Red** after $\Delta T_y$, which allows traffic to progress along direction one. As long as each traffic signal meets its local timing constraints ($\Delta T_x$ and $\Delta T_y$), the behavior at the ZTCU level still meets the traffic schedules specified in the ZTCU table. Please note that the local FSMs can switch more often, such as if no vehicle is traveling in one direction and a single car is approaching from another direction, but for "pools" of vehicles the signaling must follow the timing constraints of the ZTCU table.

Figure 8(b) presents the upper levels of the hierarchy, including the Markov Process (ATCU) and the Functional Graph (TCU) levels, and their connection to the lower Conditional Graph level (ZTCU). For an estimated number of vehicles that pass through the area, the ZTCU has the optimal activity schedules for each estimated traffic scenarios. If the timing delay $T_{area\_i}$ of a ZTCU schedule exceeds the necessary timing constraint set for the area traversal, then the traffic light controllers are adjusted so that the new traffic scenarios has a schedule that meets the constraint.

Another action set by the ATCU is to distribute the traffic flow among the two areas so that the set timing constraints are met and the total number of cars passing through the area is maximized. This decision making is achieved by setting up a Markov process for the area in which the states correspond to the traffic scenarios of the ZTCU (eight scenarios in our case study). The transition rates between the states are the probabilities $P_{(x,y,z)\rightarrow(u,v,w)}$ of shifting from one scenario to another. ATCU maintains a macro-level perspective of the traffic flow in which the moving of vehicles appears like a continuum flow without losses (the number of vehicles that enters a zone is equal to the number of vehicles that exists). For the example in Figure 7, $In_1+In_3+In_6=Out_9$ (indexes reflect the corresponding street segment). The following paragraphs introduce briefly the concept of Continuous-Time Markov Decision Processes, and explain how this formalism is used to model the higher levels of the ATCU decision making.

A Continuous-Time Markov Decision Process (CTMDP) is defined as the following sixth-tuple:
$$\{I, A, A(), q, k, r\}$$
Where $I$ is a finite state space, $A$ is a finite set of actions, $A()$ is the function indicating the actions associated to any state in $I$, $k$ is the number of criteria, $r_k(i,a)$ is the reward rate for criterion $k$, and $q(i,j,a)$ is the transition rate from state $i$ to state $j$, if action $a$ is conducted. For ATCU decision making, the different traffic scenarios define the states of a CTMDP, such one state corresponds to low traffic conditions through each intersection of the zone. Hence, one state corresponds to one column in the CTG scheduling table. If the CTMDP corresponds to multiple scheduling tables then each column generates a different state in the table. The action set $A$ represents the alternative routes that are available for moving inside a zone. The action set correlates to the available traffic scenarios and also the traffic parameters $T_i^{ex}$ and $N_i$. Rewards $r_k$ are the number of cars passing through a certain zone in a time equal to the time modeled through the scheduling table of the CTG. Thus, parameter $k$ is one in this case.

Decision making is expressed as the following optimization problem:
$$\text{Maximize} \sum \sum r(i,a) \cdot x_{i,a}$$
Such that
$$\sum q(i,a) x_{j,a} - \sum \sum q(i,j,a) \cdot x_{i,a} = 0$$
$$\sum \sum x_{i,a} = 1$$
$$x_{i,a} \geq 0$$
$$\sum \sum r_k(i,a) \cdot x_{i,a} \geq c_k$$

**Qualitative Decision Making using Fuzzy Logic.** An alternative to the previous approach can be based on hierarchical qualitative decision making, such as fuzzy logic and expert systems. Fuzzy logic computes the control outputs using a set of predefined rules that operate on a qualitative assessment of the inputs. The qualitative assessment (through membership functions) is critical as it decides the meaning of the inputs for the controller. Obviously, in a dynamic system, the meaning might change over time to reflect the insight gained during operation. Hence, the decision making process can be organized as a two level strategy, in which the upper level component uses learned patterns and dependencies to set the characteristics for decision making of the lower level, fuzzy logic controller.

Figure 10 shows the block diagram of the two-level hierarchical control system. The fuzzy logic controller ZLCU receives aggregated information from (i) the illumination sensors, aggregated by block *AG-il* (e.g., the intensity of the ambient light *i*), and (ii) the traffic sensors, aggregated by block *AG-t* (e.g., the traffic "density" *d*). ZLCU computes the command variable *u* (e.g., for the actuators of the ambient illumination system). The fuzzy controller is supervised by the upper level control system LCU that sets parameter *a*, which is used by the fuzzy logic controller to qualitatively assess the two inputs.

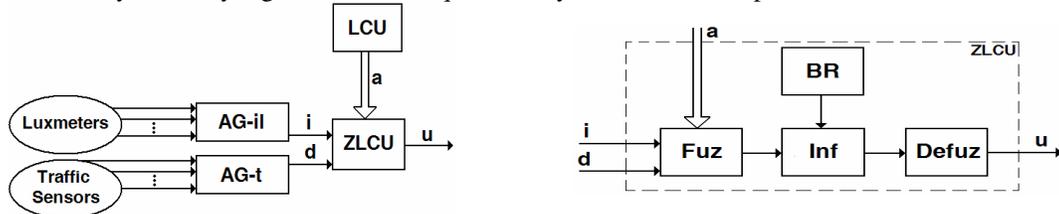

*Figure 10*: (a) Hierarchical qualitative control scheme and (b) fuzzy logic controller

Figure 10(b) presents the block diagram of the fuzzy logic controller ZLCU. The received inputs are fuzzified in the input interface block *Fuz*. Then, an inference based on the max-mi method is done on the fuzzyfied inputs using rule *BR*. The output information is defuzzified with the centroid method to produce the output *u*. The fuzzy variables $a_i$, $a_d$ and $a_u$ can take three qualitative values: small (S), medium (M), and big (B). The parameters of these membership functions on the universes of discourse are indexed with the indexes *m* for the medium value, *M* for the maximum, value and *Ml* for the limited maximum value. These parameters are modified by the high level component LCU. Samples of membership functions are given in Figure 11(a).

LCU transmits the following set of values to ZLCU:

$$a = \begin{bmatrix} a_{im} & a_{iM} & a_{iMl} \\ a_{dm} & a_{dM} & a_{dMl} \\ a_{um} & a_{uM} & a_{uMl} \end{bmatrix} \quad (1)$$

A simple rule base *BR* of the fuzzy controller is presented in Table 1.

*Table 1*: Simple rule base

| u | | d | | |
|---|---|---|---|---|
| | | S | M | B |
| | B | S | S | S |
| i | M | S | M | M |
| | S | M | B | B |

Figure 4 presents the surface of the fuzzy controller $u = f(i,d)$ for a particular value set of the parameters (0.5; 1; 1.2) on the universe of discourse [0; 1.2] for all three fuzzy variables. Changes in the membership function parameters and the rule base will change the above surface.

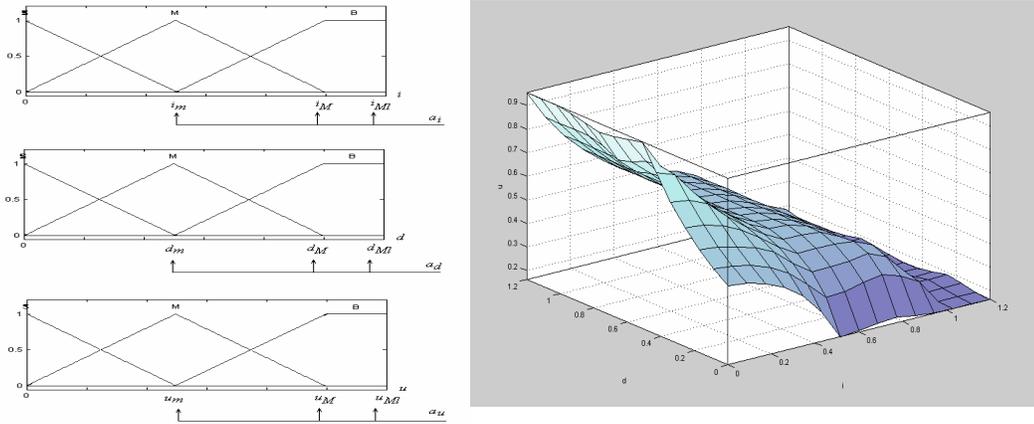

*Figure 11*: (a) Membership functions and (b) surface of the fuzzy controller

## 5. Specification for Integrated Decision Making

The paradigm discussed in Section 4 ought to support interdependent applications of different types, as the global optimization goal might not involve a certain application. For example, increasing the traffic flow during night time is linked to the illumination control as optimal illumination might allow higher vehicle speeds while still maintaining safety. This requires that the decision making (DM) modules across the entire semantic hierarchy utilize a uniform specification model, which is independent of the application type and the control strategy utilized by each individual DM module. The specification model must also allow easy decision making across the DM modules for different applications.

The specification model proposed for decision making in large scale urban environments is based on a *goal-oriented* description paradigm, in which each DM module has well-defined goals. The goals are set by the application, such as providing the required illumination intensity, or maximizing the traffic flow through a region. In addition, each DM module has limited set of capabilities, such as the maximum illumination it can provide, the highest traffic flow on a route, communication distance and bandwidth, local energy resources, and so on. Module inputs are either connected to the sensors or connections from DM modules positioned on the same or higher semantic levels. DM outputs are either actuation and control signals or connections to modules on the same or lower semantic levels. Each module conducts internally a decision making process using models similar to those detailed in Section 4, however these models are not accessible outside a DM module. Figure 12 illustrates the goal-based specification concept.

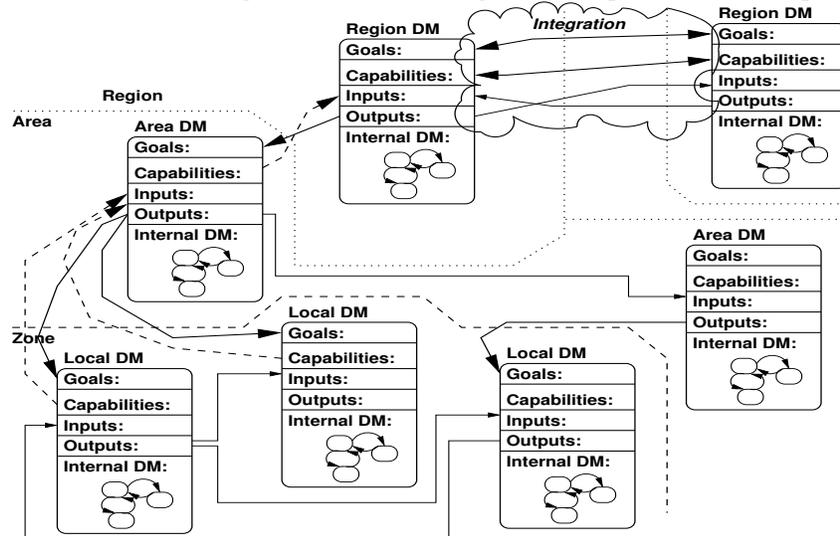

*Figure 12*: Uniform specification approach for integrated decision making

Different kinds of interactions can be set up between the participating DM modules:
- *Collaborative interactions*: Collaborative interactions are set up between DM modules on the same semantic level and which have non-conflicting (non-competing) goals, such as the individual traffic light controllers which must all offer optimal lighting intensity of an area. The DM modules interact with each other through inputs and outputs, e.g., one DM module produces outputs that serve as inputs to another module. The goals and capabilities of DM modules are not affected by these kinds of interactions.
- *Competing interactions:* Such interactions are set up between DM modules from the same semantic level but with competing goals, like maximizing the illumination of a certain zone but also minimizing the energy consumption of the zone (the latter can be set by the power grid application). Similarly to collaborative interactions, DM modules interact through their inputs and outputs but cannot change their goals or capabilities components.
- *Guiding interactions:* Guided interactions are between DM modules at consecutively higher levels in the semantic hierarchy. The modules at the upper levels generate outputs that are used to set the goals of the modules on the lower levels. This way the first modules are steering the goals and hence the behavior of the latter modules. For example, the DM modules at the area level set constraints, which then become part of the goals section of the related traffic signal controllers.
- *Enabling interactions:* Enabling interactions are information transfers from the lower semantic levels to the upper levels. A lower DM module transmits information about its capabilities to an upper DM module, so that the second uses this knowledge during a decision making process that might affect the goals set for the lower module through guided interactions. Enabling interactions is the information link through which the upper decision making modules acquire knowledge about the actions that are ongoing in the real world.

These concepts are illustrated in Figure 13 which shows the interactions and integration of the decision making mechanisms for different infrastructure management subsystems in a city, such as street lighting control, traffic control, pollution monitoring, and disaster management.

## 6. Future Research Directions

Several main research problems emerge related to the proposed decision making approach for large scale urban environments. The identified problems can be grouped into three categories, and are described next. The starting point in addressing the problems represents the developing of more orthogonal applications, like heat, water, gas, and power distribution, fast response in emergency situations, energy harvesting on large areas, environmental monitoring (including noise levels and pollution), social networking, and many more. Developing a suitable testbed for simulating these applications is also very important.

**A. Exploring the formal model: semantic hierarchy and specification**. An important goal is to study the full potential for decision making of the proposed approach based on semantic hierarchy. The paper details a hierarchy based on Finite State machines, Conditional task Graphs, Markov Decision Processes, and Functional Graphs, but other quantitative or qualitative models can be incorporated into the hierarchy too. Some of the formalisms that we plan to research include Petri Nets, Signal Flow Graphs, Timed Automata, Hybrid Automata, Flow based models, linear and nonlinear programming, and likely other too. Also, it is important to understand the interactions between the mathematical formalisms, including techniques for propagating constraints top-down from the more abstract to the more precise levels, and also bottom-up for specifying feasibility constraints at the higher levels. Moreover, it is important to understand how the propagated constraints relate to the mathematical formalisms used in the hierarchy, as the same constraints might be utilized for different specification formalisms. The scalability of the hierarchical model for different formalisms must be characterized too. Finally, the current formalism targets only static sensing nodes but it is important to extend it for mobile nodes too.

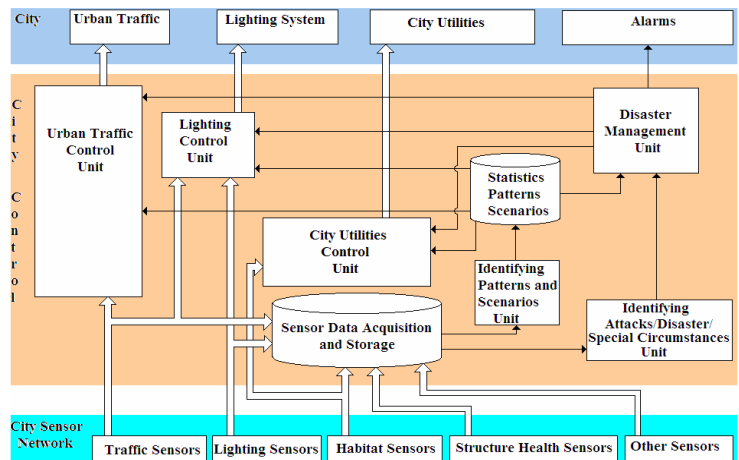

*Figure 13*: Integrated decision making in a city

Another important issue relates to lay out the mathematical foundation of collaborative interactions between decision making systems. This is important for understanding what other collaboration types are possible, if entities interact though goals, capabilities, inputs, and outputs. Moreover, the specific manifestation of collaborative interactions for different semantic models needs to be addressed too.

**B. Online decision making algorithms**. A set of specific online decision making algorithms ought to be developed for the semantic hierarchy model. This includes algorithms for forming clusters based on application goals, electing virtual leaders for each cluster, and instantiating the virtual hierarchy at the level of each embedded sensing node. This activity also implies identifying the best collaboration model among related decision modules, including finding the correlations that emerge dynamically during operation.

Also, the implications of the semantic and goal based operation mode on the underlying wireless and wired communication protocols and routing algorithms needs to be explored too.

Other important research problems refer to the actual optimization algorithms that are used for online optimization of the goals expressed in different semantic models. This includes online algorithms for resource allocation and scheduling of data flow graphs and Conditional Task Graphs, and solving equations expressed as Markov Decision Processes or flow models. Moreover, fast algorithms are needed for integrating data from the low levels into constraints for the upper semantic levels, and also for propagating top-down constraints form the upper levels to the constituent decision making entities. These algorithms ought to tackle a variety of trade-offs, like cost, speed, reliability, energy consumption, etc. Also, the importance of these constraints might also change dynamically during execution, and thus decision making should be capable to respond quickly to changes. A possibility involves developing algorithms that can adapt to new requirements by incrementally changing the current solution. Considering that the adapted solutions are not optimal, it is important to understand and estimate the performance loss of an incrementally adapted solution as compared to the optimized solution.

**C. Performance prediction for reliable operation**. A challenging objective is to provide the means for accurately estimating in real-time the performance of alternative decisions for new situations. This problem is essential for devising any efficient and reliable control algorithm. For an alternative, estimations should describe performance attributes, like quality (e.g., illumination intensity for lighting control), response time, precision, energy consumption, cost, etc. The predicted performance should include both continuous valued/time and discrete valued/time performance. Attributes like energy consumption, illumination intensity, response time, and travel time are examples of continuous valued-continuous time performance. Examples of discrete-valued attributes are the number of cars passing through a zone, the bit precision of a decision, etc. Performance estimation should rely at a minimum on simulation, instead macromodeling techniques ought to be developed based on linear and nonlinear interpolation, piecewise linear modeling, neural networks, wavelet functions, and so on. Also, it is important to detect and extract online any new patterns and correlations that emerge during operation, and use this information to further optimize the response of the decision making systems.

## 7. Current Information Systems for Large Urban Environments

A number of recent research projects focus on developing wireless sensor networks for applications in urban environments. The monitoring of the noise level in a city and building the related noise maps is presented in [Santini (2007)]. The TIME (Transport Information Monitoring Environment) project [Yoneli (2005)] targets the improving of the traffic efficiency in Cambridge UK by monitoring and processing information collected in real-time about traffic. The U-City approach [Licalzi (2005)] attempts to connect the information systems in residential, governmental, commercial, and healthcare areas. The CarTel project [Hul (2006)] focuses on distributed mobile sensor networks for monitoring traffic. The approach is based on opportunistic data muling through mobile sensors, and offers delay-tolerant pipes and delay-tolerant stack for data transport. The concept of adaptive traffic-light control through Collaborative Reinforcement Learning is introduced in [(Cunningham)]. Several other approaches are detailed next.

The CitySense concept [Murty (2007)] is wireless sensor network based infrastructure for developing application for large urban environments. Possible applications for CitySense include serving web pages, VoIP, social networking, and monitoring the networks deployed in a city. CitySense assumes static deployment of powerful computing platforms equipped with radios using the 802.11 wireless protocol to transmit over a mesh network. Hence, each node has the capabilities to perform locally complex computations and to communicate over multiple channels at high data rates. CitySense also relies on moving vehicles to provide connectivity to the areas which are hard to communicate with through radios. It uses predictions about the trajectory of the vehicle to prepare in advance the data to be communicated to the vehicle. Also this information is used to decide the aggregation points of the collected information.

The MobEyes system [Lee (2006)] has been developed for monitoring and data acquisition in urban environments. The system uses moving vehicles such as cars to transport necessary information to/from areas in which wireless communication cannot be offered. The communicated information represents a summary of the entire information harvested by the nodes during operation. The single-hop communication protocol (called MDHP) transmits the missing data summaries to a vehicle that enters the communication range of the sensor. The summaries are stored locally by the sensor or collected from its neighbors. Moving police agents collect the important summaries. The protocol is scalable, and requires acceptable large latency overhead but cannot tackle real-time constraints.

The MetroSense approach [Campbell (2006)] also tackles the problem of data gathering in urban settings, and exploits mobile vehicles for improving the covered area. This is very important for keeping the cost of the network down and also to offer connectivity if of remote zones. The authors propose an opportunistic delegation model in which tasks are temporarily re-assigned to available nodes. Similar to MobEyes, vehicles are used to carry information to remote areas. The authors propose four protocols for moving data through vehicles: lazy uploading, lazy tasking, direction-based muling, and adaptive multihop. In lazy uploading, a point-to-point connection is used to transmit data to the vehicle as long as it stays in the zone covered by the radio. The next available vehicle continues the data transfer. A similar strategy is used in lazy tasking to assign tasks (through reprogramming) to new processing devices. Direction-based muling uses predictions on vehicle moving to transfer data to the vehicles that are more likely to move towards the target of the data communication. In adaptive multihop, the characteristics of the communication, such as the task boundary, change dynamically depending on static threshold values, like the load threshold. The concept is exemplified in [Eisenman (2007)] for the practical case of data collection by a network of bicycles to improve the quality of bicycle riding.

[Kansal (2004)] propose a networking infrastructure which uses autonomous mobile wireless routers for improving the energy consumption, accuracy, data rate of communication over large areas. The infrastructure also gives connectivity to remote areas. The method computes and adjusts the trajectories for the mobile routers depending on the nature of their dynamics, e.g., random, predictable, and controlled movement. A set of adaptive control procedures are suggested to adjust the speed of the mobile routers depending on the speed and latency of data collection, and the connectivity of the network. In deciding the dynamics, the procedures exploit the correlations between the sensing speed, communication distance, data rate and energy consumption.

## 8. Conclusions

Cities are complex systems in which resources and infrastructure are continuously allocated to economy and inhabitants. A city must have sufficient decision making systems for continuously dispatching resources like water, electricity, heat, traffic, etc. in flexible, autonomous, scalable, efficient, and predictable fashion. This task is challenging as the quality of decision making is a trade-off between effectiveness, comprehensiveness, and related costs. Global decisions can be more efficient, however, they are time consuming and financially expensive. In contrast, local decisions are cheap but offer mostly local decisions. In addition, the integrated nature of modern applications requires control strategies that can comprehensively tackle correlated applications or which might become correlated in certain conditions. Intriguing new control solutions can be devised based on modern technology involving ubiquitous data collection through sensors, automated analysis and prognosis, and online optimization. This enables global decision making over large areas while keeping the costs low. Also, new mechanisms can be devised to integrate and co-optimize the correlated applications. This expands the optimization space, and increases the reliability of decision making in new situations.

This paper proposes a new decision making paradigm for optimizing the continuous, real-time allocation of resources to satisfy demands in large cities. The paper refers to street lighting and traffic control as case studies to illustrate the paradigm. The proposed concept is structured as a *semantic hierarchy* in which different decision making models and strategies coexist and interact to produce flexible, autonomous, scalable, efficient, and predictable decisions. For scalability, only a reasonably large number of modules collaborate with each other to reach global decisions. Flexibility and autonomy results by using reactive models for the bottom decision making levels. Efficiency and predictability is due to the reactive behavior being constrained by the upper semantic levels, which use more deterministic models. The upper levels compute the limits within which reactive decision making must operate, so that the overall goals and constraints are not violated. The data sampled from the environment is aggregated and propagated to the upper semantic levels, where it is used for global decision making. Finally, the paper discusses the defining elements of the specification formalism for integrating different but related applications. Interactions between decision modules, such as collaboration, competition, guidance, and enabling interactions, are defined based on the goals, capabilities, inputs, and outputs of each module.

While the proposed decision making paradigm is applicable to other theoretical formalisms too, this paper refers to a semantic hierarchy based on Finite State Machines (FSMs), Conditional Task Graphs (CTGs), Continuous Time Markov Decision Processes (CTMDPs), and Functional Graphs (FGs). The hierarchy is exemplified for coordinated traffic signal control, a main application in modern cities. Decisions at successively higher semantic levels are used to cover increasingly broader geographical areas: FSMs implement reactive control guided by signals coming from sensors as well as neighboring FSMs, CTGs schedule the related activities over time for different traffic conditions, CTMDP use macroscopic descriptions of the system to conduct scenarios-specific optimizations, and FGs regulate the global allocation of resources and infrastructure to global demands.

Three main research problems emerge related to the proposed decision making approach. First, it is important to study semantic hierarchies based also on other formalisms like Petri Nets, Signal Flow Graphs, Timed Automata, Hybrid Automata, Flow based models, linear and nonlinear programming, and so on. Moreover, it is needed to devise the theoretical foundations for propagating constraints top-down from the more abstract to the more precise levels, and also bottom-up for specifying feasibility constraints at the higher levels. Second, efficient online decision making algorithms ought to be developed for the semantic hierarchy model, e.g., algorithms for forming goal oriented clusters, electing virtual leaders, and mapping the virtual hierarchy onto each embedded node. Also, new algorithms need to be designed for resource allocation and scheduling at different levels of the model hierarchy while tackling trade-offs, like cost, speed, reliability, energy consumption, etc. As the importance of these constraints might change during execution, the algorithms should incrementally change the current solution to adapt it to the new requirements. Third, procedures for accurate, real-time performance estimation ought to be devised. The tackled performance should include both continuous valued/time and discrete valued/time performance. Moreover, performance estimation should rely mostly on macromodeling, so that it can be performed in real-time. Other tasks relate to detecting and extracting online any emerging patterns and correlations, and using this information to further optimize decision making.